\documentclass{elsarticle}

\usepackage{hyperref}
\usepackage{lineno}
\usepackage[ruled]{algorithm2e}
\usepackage[utf8]{inputenc}
\usepackage[T1]{fontenc}
\usepackage{lmodern}
\usepackage{fancyhdr}
\usepackage{graphicx}
\usepackage{mathrsfs}
\usepackage{amsmath}
\usepackage{amsfonts}
\usepackage{supertabular}
\usepackage{sublabel}
\usepackage{xkeyval}
\usepackage{multimedia}
\usepackage{xcolor}
\usepackage{amssymb} 
\usepackage{upgreek} 
\usepackage{mathtools}

\SetAlFnt{\algofont}
\SetAlCapFnt{\algofont}
\SetAlCapNameFnt{\algofont}
\SetAlCapHSkip{0pt}
\IncMargin{-\parindent}

\journal{ArXiv}

\begin{document}

\begin{frontmatter}

\title{Trivergence of Probability Distributions, at glance}
\tnotetext[mytitlenote]{LIA/Université d'Avignon et des Pays de Vaucluse}


\author[mymainaddress,otra]{Juan-Manuel Torres-Moreno\corref{mycorrespondingauthor}}
\cortext[mycorrespondingauthor]{Corresponding author}
\ead{juan-manuel.torres@univ-avignon.fr}
\ead[url]{lia.univ-avignon.fr/chercheurs/torres}

\address[mymainaddress]{Laboratoire Informatique d'Avignon/UAPV BP 91228, 84911 France}
\address[otra]{Ecole Polytechnique de Montr\'eal, Montréal (Québec) Canada}

\begin{abstract}
In this paper we introduce the intuitive notion of trivergence of probability distributions (TPD). 
This notion allow us to calculate the similarity among triplets of objects. 
For this computation, we can use the well known measures of probability divergences like Kullback-Leibler and Jensen-Shannon.
Divergence measures may be used in Information Retrieval tasks as Automatic Text Summarization, Text Classification, among many others.
\end{abstract}


\end{frontmatter}


\section{Introduction}

A statistical distance defines a measure of distance between two objects.
This measure of distance may be interpreted as a distance among two probability distributions of two populations.
Moreover, a metric is a measure defined on a set $\cal X$ as a function $d$ such as, $\forall x,y \in \cal X$,
$d : \cal{X} \times \cal{X} \mapsto R+$. 
$d$ respects the following conditions:
\begin{enumerate}[i)]
\item $d(x, y) \ge 0$ 
\item $d(x, y) = 0   \textrm{ iff }   x = y$
\item $d(x, y) = d(y, x)$ 
\item $d(x, z) \le d(x, y) + d(y, z)  $
\end{enumerate}

Several measures of distance are not considered as metrics because they do not fulfill one or more of these conditions.
These measures are known as divergences.
This is the case of Kullback-Leibler divergence $D_{\mathrm{KL}}$, that in particular, violates the conditions ii) and iii).
In other hands, the Jensen-Shannon divergence $D_{\mathrm{JS}}$ is a metric. It corresponds to the symmetrical version of the  $D_{\mathrm{KL}}$ divergence. 

In this paper we introduce the notion of distance among three objects as a trivergence $\uptau$ of probability distribution.
The main idea is based on intuitive properties of divergences.

The rest of the paper is organized as follows:
in Section §\ref{sec:div} we outline the divergences using probability distributions and smoothing.
Section §\ref{sec:triv} introduces the preliminaries of notion of trivergence.
Sections §\ref{sec:dem:KL} and §\ref{sec:dem:JS} compute the trivergence
as a product of divergences and as a compound divergence function.
Finally Section §\ref{sec:conc} shows the discussion and the conclusions. 

\section{Preliminaries: divergences of probability distributions with smoo\-thing}
\label{sec:div}

In the follows, we recapitulate the divergence functions of probability distributions:
the  Kullback-Leibler divergence \cite{cover1991elements} and the  Jensen-Sha\-nnon symmetrical divergence~\cite{EndSch03}.  

\subsection{Kullback-Leibler divergence}

The divergence of Kullback-Leibler or relative entropy is a distance between two probability distributions $p$ and $q$ is defined by the equation: 
\begin{equation}
\label{KL}
	D_{\mathrm{KL}}(p\Vert q)=\displaystyle \sum_{w\in p}p_w\log\frac{p_w}{q_w}
\end{equation}
The logarithm is in base~2, but we adopted the notation convention $\log_2$ as $\log$.

Of course, $q_w=0$ for a few items $w$, because not all items of $p$ are in $q$. 
In this case, expressions like $p\log\frac{p}{0} \rightarrow \infty$ may occur if $q_w=0$, i.e. when the item $w \notin q$ (see by example the Figure \ref{fig:conj}).
To avoid this situation, in an empirical way, a smoothing process is used for estimating the probability of unseen items. 
In the literature there are several smoothing techniques, for example Good-Turing, Back-Off, etc. \cite{Chen98anempirical,manning:99}.
In this paper, we will use a very elementary smoothing:
\begin{equation}
\label{eq:smooth}
q_w = 
 \begin{dcases}
  ~\frac{C_w^q}{|q|} & \textrm{ if } w \in q \\
  ~\frac{1}{|T|} & \textrm{ elsewhere}
 \end{dcases}
\end{equation}

\noindent where $p$ and $q$ are the probability distributions, $p_w=\frac{C^{p}_{w}}{\vert p\vert}$, $q_w$ is defined by equation (\ref{eq:smooth}), $C^p_w$ is the number of ocurrences of the item $w \in p$, $C^q_w$ is the number of ocurrences of the item $w \in q$, $|p|=$ total number of distinct items $\in p$, $|q|=$ total number of distinct items $\in q$ and $\vert T\vert=\vert p\vert + \vert q\vert$. 
In other hands, we assume that $|p| > |q|$, then the divergence is calculated from $p$ to $q$.

The Kullback-Leibler distance is not a metric in a mathematical sense, because despite meeting that $D_{\mathrm{KL}}(p\Vert q)\geq 0$ with equality if and only if $p=q$, it is not symmetrical and it does not respect the triangle inequality.

\subsection{Jensen-Shannon divergence}

The Jensen-Shannon divergence\cite{EndSch03} or symmetrical distance of Kullback-Leibler between two probability distributions $p$ and $q$ over the same alphabet $\cal{X}$ is defined by the equation: 
\begin{equation}
\label{JS} 
  D_{\mathrm{JS}}(p\Vert q)=\frac{1}{2} \left\{\sum_{w\in \cal X}p_w\log\frac{2 p_w}{p_w + q_w} + \sum_{w\in \cal X}q_w\log\frac{2 q_w}{p_w+q_w}\right\}
\end{equation}
\noindent with the same conventions for $p$, $q$, $\vert p\vert$, $\vert q\vert$, $\vert T\vert$, $p_w$, $q_w$, $C^p_w$ and $C^q_w$ as in equation (\ref{KL}); and the same elementary smoothing (\ref{eq:smooth}). 
The logarithm is also in base 2, but we adopted the same convention for $\log_2$. 
$\sqrt{D_{\mathrm{JS}}}$ is a metric in a mathematical sense.

\section{Trivergence of probability distributions}
\label{sec:triv}

\begin{figure}[h!]
  	\centering
  	\includegraphics[scale=0.25]{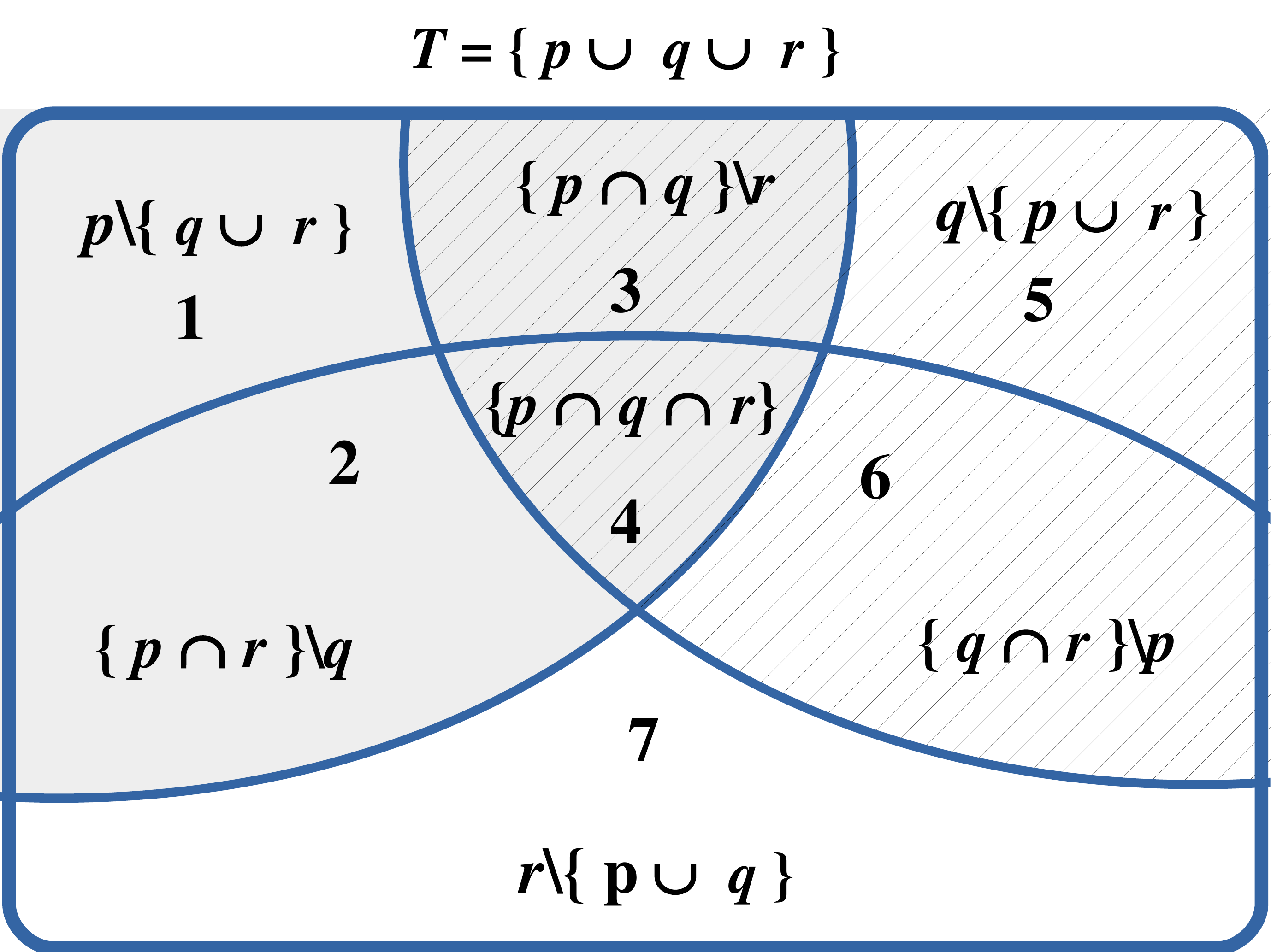}
  	\caption{Decomposition of the distributions $p$, $q$ and $r$ in subsets.}
  \label{fig:conj}
  \end{figure}

In order to define the trivergence between three probability distributions we will use divergence measures.
Let $p$, $q$ and $r$ be three probability distributions and $T= \{p \cup q \cup r\}$, with cardinality $|T|$.
Figure \ref{fig:conj} shows the partitioning of the $T$ set in 7 regions. 

We defined two ways to calculate the trivergence $\uptau$, as a product of divergences and as a compound divergence function:
\begin{enumerate}[i)]
\item 
Product of divergences:
\begin{equation}
\label{eq:prod0}
\uptau^\pi(p || q || r) = 
\begin{cases} 
	D(p||q) \cdot  D(q||r) \cdot D(p||r) ; \\
	D(q||p) \cdot  D(r||q) \cdot D(r||p) \\
\end{cases}
\end{equation}
\item Compound divergence function: 
\begin{equation}
\label{eq:comp0}
\uptau^c(p || q || r) = 
\begin{cases}
	D[~ p|| D(q||r) ~] &;~ D[~ p|| D(r||q) ~];  \\
	D[~ q|| D(p||r) ~] &;~ D[~ q|| D(r||p) ~];\\
	D[~ r|| D(p||q) ~] &;~ D[~ r|| D(q||p) ~]; \\ 
	D[~ D(q||r)||p ~] &;~ D[~ D(r||q) || p~]; \\
	D[~ D(p||r)||q ~] &;~ D[~ D(r||p)||q ~]; \\
	D[~ D(p||q)||r ~] &;~ D[~ D(q||p)||r ~]
\end{cases}
\end{equation}
\end{enumerate}
In both cases, if we use the following restriction: 
\[|p|>|q|>|r|\]
the definition of trivergence is, in particular, sorted by their cardinality. 
Then, we have for the product:
\begin{equation}
\uptau^\pi(p || q || r) = 
	D(p||q) \cdot  D(q||r) \cdot D(p||r) 
\end{equation}
And for the compound function: 
\begin{equation}
\uptau^c(p || q || r) = D[~ p|| D(q||r) ~]
\end{equation}

In order to clarify the weight of the smoothing (equation \ref{eq:smooth}) for $p_w, q_w$ and $r_w$, from Figure \ref{fig:conj} we have for each region that:
\begin{enumerate}
\item \{$p \backslash \{q \cup r$\}\}: $q_w = r_w = 0$;  
\item \{$p \cap r \} \backslash q$: $q_w = 0$;
\item \{$p \cap q \} \backslash r$: $r_w = 0$;  
\item \{$p \cap q \cap r\}$: $p_w \ne 0, q_w \ne 0, r_w \ne 0$;  
\item \{$q \backslash \{p \cup r$\}\}: $p_w = r_w = 0$;
\item \{$q \cap r \} \backslash p$: $p_w = 0$; 
\item \{$r \backslash \{p \cup q$\}\}: $p_w = q_w = 0$. 
\end{enumerate}

In this paper, we will use both Kullback-Leibler $D_{\mathrm{KL}}$ \cite{cover1991elements} and Jensen-Shannon $D_{\mathrm{JS}}$~\cite{EndSch03} divergences in order to calculate the trivergence $\uptau^{\pi,c}(p||q||r)$.

\section{Distribution using Kullback-Leibler divergence}
\label{sec:dem:KL}

\subsection{$\uptau^\pi$ as product of KL divergences}

\noindent \textbf{Definition.} 
 Let $p$, $q$ and $r$ be three probability distributions where \[|p|>|q|>|r|\]
 and $T= \{p \cup q \cup r\}$, with cardinality $|T|$.
 The Kullback-Leibler trivergence  between $p$, $q$ and $r$, sorted by their cardinality is defined as a product of divergences:  
\begin{equation*}
 \uptau^\pi_{\mathrm{KL}}(p || q || r) = D_{\mathrm{KL}}(p||q) \cdot  D_{\mathrm{KL}}(q||r) \cdot D_{\mathrm{KL}}(p||r)
\end{equation*}

Calculating simultaneously for $p$, $q$ and $r$:
\begin{equation}
\label{eq:KLpq}
 D_{\mathrm{KL}}(p ||q) = \sum_{x\in p}p_w\log\frac{p_w}{q_w} 
\end{equation}
\begin{equation}
\label{eq:KLqr}
 D_{\mathrm{KL}}(q ||r) = \sum_{x\in q}q_w\log\frac{q_w}{r_w} 
\end{equation}
\begin{equation}
\label{eq:KLpr}
 D_{\mathrm{KL}}(p ||r) = \sum_{x\in p}p_w\log\frac{p_w}{r_w} 
\end{equation}

\noindent From the equation (\ref{eq:KLpq}):   
\begin{eqnarray}
\label{eq:KLpq1}
  \sum_{x\in p}p_w\log\frac{p_w}{q_w}  &=&  
  \sum_{x\in p \backslash q}p_w\log\frac{p_w}{q_w} + \sum_{x\in p \cap q}p_w\log\frac{p_w}{q_w}
\end{eqnarray}
\noindent and using the smoothing from the equation (\ref{eq:smooth}): 
\begin{equation*}
\label{eq:KLpq2}
 \sum_{x\in p } p_w\log\frac{p_w}{q_w} = 
\begin{dcases}
  \sum_{x\in p \backslash q}  \frac{C_w^p}{|p|} \log \frac{|T|C_w^p}{|p|} & \textrm{smooth } q_w=\frac{1}{|T|} \\
  \sum_{x\in p \cap  q} \frac{C_w^p}{|p|} \log \frac{|q|}{|p|} \frac{C_w^p}{C_w^q} & \textrm{without smooth}
\end{dcases}
\end{equation*}

\noindent From the equation (\ref{eq:KLqr}):   
\begin{eqnarray}
\label{eq:KLqr1}
  \sum_{x\in q}q_w\log\frac{q_w}{r_w}  &=&  
  \sum_{x\in q \backslash r} q_w\log\frac{q_w}{r_w} + \sum_{x\in q \cap r} q_w\log\frac{q_w}{r_w} 
\end{eqnarray}
\noindent and using the smoothing from the equation (\ref{eq:smooth}): 
\begin{equation*}
\label{eq:KLqr2}
  \sum_{x\in q } q_w\log\frac{q_w}{r_w}  = 
\begin{dcases}
  \sum_{x\in q \backslash r} \frac{C_w^q}{|q|} \log \frac{|T|C_w^q}{|q|}  & \textrm{smooth }  r_w = \frac{1}{|T|} \\ 
  \sum_{x\in \{q \cap r\} } \frac{C_w^q}{|q|} \log \frac{|r|}{|q|} \frac{C_w^q}{C_w^r} & \textrm{without smooth}
\end{dcases}
\end{equation*}

\noindent From the equation (\ref{eq:KLpr}):   
\begin{eqnarray}
\label{eq:KLpq1}
  \sum_{x\in p}p_w\log\frac{p_w}{r_w}  &=&  
   \sum_{x\in p \backslash r} p_w\log\frac{p_w}{r_w} + \sum_{x\in p \cap r} p_w\log\frac{p_w}{r_w} 
\end{eqnarray}
\noindent and using the smoothing from the equation (\ref{eq:smooth}): 
\begin{equation*}
\label{eq:KLpr1}
  \sum_{x\in p } p_w\log\frac{p_w}{r_w} =
\begin{dcases}
  \sum_{x\in p \backslash r} \frac{C_w^p}{|p|} \log \frac{|T|C_w^p}{|p|}               & \textrm{smooth } r_w=\frac{1}{|T|} \\ 
  \sum_{x\in \{p \cap r\} } \frac{C_w^p}{|p|} \log \frac{|r|}{|p|} \frac{C_w^p}{C_w^r} & \textrm{without smooth}
\end{dcases}
\end{equation*}

\noindent therefore:
\begin{equation}
\label{eq:KLpqsimp}
 D_{\mathrm{KL}}(p||q) =
 \sum_{x\in p\backslash q} \frac{C_w^p}{|p|} \log \frac{|T|C_w^p}{|p|} +
  \sum_{x\in \{p \cap q \}} \frac{C_w^p}{|p|} \log \frac{|q|}{|p|} \frac{C_w^p}{C_w^q}
\end{equation}

\begin{equation}
\label{eq:KLqrsimp}
 D_{\mathrm{KL}}(q||r) =
  \sum_{x\in q\backslash r} \frac{C_w^q}{|q|} \log \frac{|T|C_w^q}{|q|} +
  \sum_{x\in \{q \cap r\}} \frac{C_w^q}{|q|} \log \frac{|r|}{|q|} \frac{C_w^q}{C_w^r}
\end{equation}

\begin{eqnarray}
\label{eq:KLprf}
 D_{\mathrm{KL}}(p||r) =
  \sum_{x\in p \backslash r} \frac{C_w^p}{|p|} \log \frac{|T|C_w^p}{|p|} +
  \sum_{x\in \{p \cap r\}}  \frac{C_w^p}{|p|} \log \frac{|r|}{|p|} \frac{C_w^p}{C_w^r} 
\end{eqnarray}

\subsection{$\uptau^\pi$ as compound divergence function}
\label{sec:dem:KL1}

\noindent \textbf{Definition} 
 Let $p$, $q$ and $r$ be three probability distributions where \[|p|>|q|>|r|\]
 and $T= \{p \cup q \cup r\}$, with cardinality $|T|$.
 The Kullback-Leibler trivergence  between $p$, $q$ and $r$, sorted by their cardinality is defined as a compound divergence function: 
\begin{equation*}
 \uptau^c_{\mathrm{KL}}(p || q || r) = D_{\mathrm{KL}}\left[~ p ~||~ \frac{D_{\mathrm{KL}}( q || r)}{|q|}~ \right]
\end{equation*}
We computed $\frac{D_{\mathrm{KL}}( q || r)}{|q|}$ in order to consider this fraction suchs as a probability. 

Firstly, we calculate:
\begin{eqnarray*}
\label{eq:KL1}
 D_{\mathrm{KL}}(q ||r) &=& \sum_{w \in q}q_w\log\frac{q_w}{r_w} 
\end{eqnarray*}
however
$
\sum_{w\in q}q_w\log\frac{q_w}{r_w} 
$
is defined by equation (\ref{eq:KLqrsimp}), therefore using a smoothing in the case of unseen events:
\begin{equation}
\label{eq:KL3}
  \uptau^c_{\mathrm{KL}}(p || q || r) =  
  \begin{dcases}
	\sum_{x \in p \cap q} p_x\log\frac{|q|p_x}{D_{\mathrm{KL}}(q||r)}\\
	\sum_{x \in p \backslash q} p_x\log |T| p_x & \textrm{if } {D_{\mathrm{KL}}(q||r)}=0;
  \end{dcases}
\end{equation}

\section{Distribution using Jensen-Shannon divergence}
\label{sec:dem:JS}

\subsection{$\uptau^\pi$ as product of JS divergences}

\noindent \textbf{Definition.} 
 Let $p$, $q$ and $r$ be three probability distributions where \[|p|>|q|>|r|\]
 and $T= \{p \cup q \cup r\}$, with cardinality $|T|$.
 The Jensen-Shannon trivergence between $p$, $q$ and $r$, sorted by their cardinality is defined as a product of divergences:
\begin{equation*}
 \uptau^\pi_{\mathrm{JS}}(p || q || r) = D_{\mathrm{JS}}(p||q) \cdot  D_{\mathrm{JS}}(q||r) \cdot D_{\mathrm{JS}}(p||r)
\end{equation*}

We defined:
\[
P^{pq}_w = p_w\log\frac{2p_w}{p_w+q_w} ; Q^{pq}_w= q_w\log\frac{2q_w}{p_w+q_w} 
\]
\[
Q^{qr}_w = q_w\log\frac{2q_w}{q_w+r_w} ; R^{qr}_w= r_w\log\frac{2r_w}{q_w+r_w}
\]
\[
R^{pr}_w = r_w\log\frac{2r_w}{r_w+p_w} ; P^{pr}_w= p_w\log\frac{2p_w}{r_w+p_w}
\]

Calculating simultaneously for $p$, $q$ and $r$:
\begin{eqnarray}
\label{eq:JSpqr}
 D_{\mathrm{JS}}(p ||q) &=& \frac{1}{2}\sum_{w\in \{p\cup q\}}\left\{ P^{pq}_w + Q^{pq}_w \right \}\\
 D_{\mathrm{JS}}(q ||r) &=& \frac{1}{2}\sum_{w\in \{q\cup r\}}\left\{ Q^{qr}_w + R^{qr}_w \right \}\\
 D_{\mathrm{JS}}(p ||r) &=& \frac{1}{2}\sum_{w\in \{p\cup r\}}\left\{ P^{pr}_w + R^{pr}_w \right \}
\end{eqnarray}

For $2D_{\mathrm{JS}}(p ||q)$ we have:
\begin{eqnarray*}
    \sum_{w\in p\cup q}\left\{ P^{pq}_w + Q^{pq}_w \right \} 
&=& \sum_{w\in p \backslash q} P^{pq}_w + Q^{pq}_w + \sum_{w\in p\cap q } P^{pq}_w + Q^{pq}_w \nonumber \\
&+& \sum_{w\in q \backslash p } P^{pq}_w + Q^{pq}_w 
\end{eqnarray*}

\noindent and using the smoothing for $p_w$ and $q_w$ from the equation (\ref{eq:smooth}): 
\begin{equation}
\sum_{w\in p\cup q}  P^{pq}_w + Q^{pq}_w  =
\begin{dcases}
\sum_{w\in p \backslash q} \frac{C_w^p}{|p|} \log\frac{2|T|C_w^p}{|T|C_w^p+|p| } + \frac{1}{T}\log\frac{2|p|}{|T|C_w^p+|p|} ; 
 \; q_w = \frac{1}{|T|} \\ 
\sum_{w\in p\cap r } \frac{C_w^p}{|p|}\log\frac{2|q|C_w^p}{ |q|C_w^p  + |p|C_w^q } + \frac{C_w^q}{|q|}\log \frac{2|p|C_w^q}{|q|C_w^p+|p|C_w^q}\\ 
\sum_{w\in q \backslash p } \frac{1}{T}\log\frac{2|q|}{|T| C_w^q + |q|} + \frac{C_w^q}{|q|}\log\frac{2|T|C_w^q}{|T|C_w^q + |q| } ; \; p_w = \frac{1}{|T|}
\end{dcases}
\end{equation}

For $2D_{\mathrm{JS}}(q ||r)$ we have:
\begin{eqnarray*}
\sum_{w\in q\cup r}\left\{ Q^{qr}_w + R^{qr}_w \right \} &=& 
\sum_{w\in q \backslash r} Q^{qr}_w + R^{qr}_w + 
\sum_{w\in q\cap r }       Q^{qr}_w + R^{qr}_w  \nonumber \\
&+&
\sum_{w\in r \backslash q } Q^{qr}_w + R^{qr}_w 
\end{eqnarray*}

\noindent and using the smoothing for $q_w$ and $r_w$ from the equation (\ref{eq:smooth}):
\begin{eqnarray}
\label{eq:JSqrfinal}
\sum_{w\in q \cup r } Q^{qr}_w + R^{qr}_w = 
\begin{dcases}
\sum_{w\in q \backslash r } \frac{C_w^q}{|q|} \log\frac{2|T|C_w^q}{|T|C_w^q+|q| } + \frac{1}{T}\log\frac{2|q|}{|T|C_w^q+|q|} ; 
 \; r_w = \frac{1}{|T|} \\
\sum_{w\in q\cap r } \frac{C_w^q}{|q|}\log\frac{2|r|C_w^q}{ |r|C_w^q  + |q|C_w^r } + \frac{C_w^r}{|r|}\log \frac{2|q|C_w^r}{|r|C_w^q+|q|C_w^r}\\
\sum_{w\in r \backslash q } \frac{1}{T}\log\frac{2|r|}{|T| C_w^r + |r|} + \frac{C_w^r}{|r|}\log\frac{2|T|C_w^r}{|T|C_w^r + |r| } ; \; q_w = \frac{1}{|T|}
\end{dcases}
\end{eqnarray}

Finally, for $2D_{\mathrm{JS}}(p ||r)$ we have:
\begin{eqnarray*}
\sum_{w\in p\cup r}\left\{ P^{pr}_w + R^{pr}_w \right \} &=& 
\sum_{w\in p \backslash r} P^{pr}_w + R^{pr}_w + 
\sum_{w\in p\cap r }       P^{pr}_w + R^{pr}_w  \nonumber \\
&+&\sum_{w\in r \backslash p } P^{pr}_w + R^{pr}_w 
\end{eqnarray*}

\noindent Using the smoothing for $p_w$ and $r_w$ from the equation (\ref{eq:smooth}): 
\begin{eqnarray}
\label{eq:JSpqr2}
\sum_{w\in p \cup r} P^{pr}_w + R^{pr}_w = 
\begin{dcases}
\sum_{w\in p \backslash r} \frac{C_w^p}{|p|} \log\frac{2|T|C_w^p}{|T|C_w^p+|p| } + \frac{1}{T}\log\frac{2|p|}{|T|C_w^p+|p|} ; 
 \; r_w = \frac{1}{|T|} \\
\sum_{w\in p\cap r } \frac{C_w^p}{|p|}\log\frac{2|r|C_w^p}{ |r|C_w^p  + |p|C_w^r } + \frac{C_w^r}{|q|}\log \frac{2|p|C_w^q}{|r|C_w^p+|p|C_w^r}\\ 
\sum_{w\in r \backslash p } \frac{1}{T}\log\frac{2|r|}{|T| C_w^r + |r|} + \frac{C_w^r}{|r|}\log\frac{2|T|C_w^r}{|T|C_w^r + |r| } ; \; p_w = \frac{1}{|T|}
\end{dcases}
\end{eqnarray}

\subsection{$\uptau^c$ as compound divergence function}

\noindent \textbf{Definition.} 
 Let $p$, $q$ and $r$ be three probability distributions where \[|p|>|q|>|r|\] 
 $T= \{p \cup q \cup r\}$, with cardinality $|T|$ and $QR= \{q \cup r\}$, with cardinality $|QR|$.
 The Jensen-Shannon trivergence sorted by their cardinality, between $p$, $q$ and $r$ is defined as a compound divergence function:

\begin{equation*}
 \uptau^c_{\mathrm{JS}}(p || q || r) = D_{\mathrm{JS}}\left[~ p ~||~ \frac{D_{\mathrm{JS}} ( q || r)}{|QR|}~\right]
\end{equation*}
We computed $\frac{D_{\mathrm{JS}}( q || r)}{|q| + |r|}$ in order to consider this fraction suchs as a probability. 

First, we calculate:
\begin{eqnarray*}
\label{eq:JS1}
 D_{\mathrm{JS}}(q || r) = \frac{1}{2} \left\{\sum_{w\in q\cup r}q_w\log\frac{2 q_w}{q_w + r_w} +
 \sum_{w \in q\cup r}r_w\log\frac{2 r_w}{q_w + r_w} \right\}
\end{eqnarray*}
\noindent neverthless $D_{\mathrm{JS}}(q || r)$ is defined by equation (\ref{eq:JSqrfinal}), therefore using a smoothing in the case of unseen events:
\[
	p_x =  \frac{D_{\mathrm{JS}}(q||r)}{|q| + |r|} = \frac{1}{|T|}
\]
\begin{eqnarray}
\label{eq:JS3}
   \uptau^c_{\mathrm{JS}}(p || q || r) = \frac{1}{2} \times \hspace{9.5cm} \nonumber \\
 \begin{dcases}
   \sum_{x \in p\cap \{q \cup r\}} p_x\log\frac{2 |QR| p_x}{|QR| p_x + D_{\mathrm{JS}}(q||r)} +
   \frac{D_{\mathrm{JS}}(q||r)}{|QR|} \log\frac{2  D_{\mathrm{JS}}(q||r)}{|QR| p_x + D_{\mathrm{JS}}(q||r)}  \nonumber \\
 \sum_{x \in p \backslash \{q \cup r\}} p_x\log\frac{2 |T| p_x}{|T| p_x + 1} +
 \frac{1}{|T|} \log\frac{2}{|T| p_x + 1} \hspace{1.5cm} \; ; \textrm{if } D_{\mathrm{JS}}(q||r) = 0  \nonumber \\
 \sum_{x \in \{q \cup r\} \backslash p}\frac{1}{|T|}\log\frac{2 |QR|}{|QR| + |T|D_{\mathrm{JS}}(q||r)} +
 \frac{D_{\mathrm{JS}}(q||r)}{|QR|} \log\frac{2 |T| D_{\mathrm{JS}}(q||r)}{|QR| + |T|D_{\mathrm{JS}}(q||r)} \nonumber \\
	\hspace{8.5cm} \; ; \textrm{if } p_x=0
 \end{dcases}
\end{eqnarray}

\section{Conclusions}
 \label{sec:conc}

The main contribution of this paper is the formalisation of the definition of smoothed Trivergence of Probability Distributions (TPD). 
The trivergence of three objects represented as probability distributions, was calculated using elementary functions of divergence (KL and JS). 
We have proposed two ways to compute the smoothed TPD. 
The first one uses a product of divergences and the second one uses a compound divergence function.
Divergences measures hase been used in Automatic Text Summarization \cite{louis:13,torres2010summary,saggion2010multilingual} tasks among many others.

\section*{References}

\bibliography{evaluating}

\end{document}